\documentclass[a4paper,11pt]{article}
\pdfoutput=1 
\usepackage{jheppub}

\title{AGT relation in the light asymptotic limit}
\author{Naofumi Hama}
\author{and Kazuo Hosomichi}

\affiliation{Yukawa Institute for Theoretical Physics\\Kyoto University\\Kyoto 606-8502, Japan}

\emailAdd{hama@yukawa.kyoto-u.ac.jp}
\emailAdd{hosomiti@yukawa.kyoto-u.ac.jp}

\abstract{
It is known that the path integral of correlators in Liouville theory reduces to a finite dimensional integral in the limit of vanishing coupling $b$. We take the example of four-point functions on sphere and investigate how the simple integral expression is reproduced from the path integral of gauge theory on extremely squashed ellipsoids. The simplified form for correlators suggests there is a 2D gauge theory describing the limit.
}

\keywords{Supersymmetric gauge theory, Liouville theory}

\begin{document}
\begin{flushright}
YITP-13-70
\end{flushright}
\maketitle
\section{Introduction}
\label{sec:intro}

Since the discovery by Alday, Gaiotto and Tachikawa, many nontrivial
correspondences have been found between 4D ${\cal N}=2$ superconformal
gauge theories and 2D Liouville or Toda CFTs, which can be regarded as
two different descriptions of the same wrapped
M5-branes \cite{Alday:2009aq,Wyllard:2009hg}. The basic relations are
that the instanton partition functions
\cite{Nekrasov:2002qd} correspond to conformal blocks, namely the
solutions of holomorphic Ward identities in 2D CFTs, and that the
partition functions on four-sphere \cite{Pestun:2007rz} or
ellipsoids \cite{Hama:2012bg} correspond to 2D correlation functions.
The discovery of this relation led to an extensive study of many
different aspects of wrapped M5-branes and the 6D $(2,0)$ theories,
and also brought important developments in the application of
localization technique to supersymmetric gauge theories.

AGT relation is interesting in its own right, and many proofs have been
proposed based on different ideas. For example, there is an approach
using matrix models and topological string \cite{Dijkgraaf:2009pc}, the
properties of conformal blocks in the series expanded form
\cite{Fateev:2009aw,Alba:2010qc}, the action of conformal or W
symmetries on instanton moduli spaces
\cite{Schiffmann:2012aa,Maulik:2012wi}, the reduction of gauge
theories to the theory of flat connections on Riemann surfaces
\cite{Nekrasov:2010ka,Vartanov:2013ima} or M-theory compactifications \cite{Tan:2013tq}. They all shed light on different
structures in 4D gauge theories that we have not been fully aware of.
A more direct proof would be to show that the $(2,0)$ theory
partially compactified on the Omega background
$\mathbb R^4_{\epsilon_1,\epsilon_2 }$ or 4D ellipsoid gives rise to the
Liouville or Toda CFTs, though it would be extremely difficult without
Lagrangian description of the $(2,0)$ theories at hand. For other
versions of the AGT-like relation, there have been a recent progress
regarding the compactification of $(2,0)$ theories on $S^3\times S^1$
\cite{Kawano:2012up,Fukuda:2012jr} or $S^3$
\cite{Cordova:2013cea,Lee:2013ida,Yagi:2013fda}.

In both Liouville CFT and 4D ${\cal N}=2$ SUSY gauge theories, it is
understood to a great extent how to perform (or make use of) the
explicit path integral to study certain class of observables. So it is
natural to ask whether the AGT relation can be promoted from the
correspondence between observables to a correspondence between two
system of path integrals, and to what extent. To address such a
question, one approach will be to focus first on the special cases in
which the path integration is known to simplify. As an example,
in this article we consider the so-called ``light asymptotic limit'' of
Liouville theory, where the Liouville coupling $b$ is very small and
only the vertex operators with small Liouville momenta are inserted.
We will consider the four-point function on the sphere
$S^2$, for which the Liouville path integral is known to reduce
to an integral over a hyperboloid $H_3^+$. In the $b\to 0$ limit, the
corresponding 4D gauge theory, namely the ${\cal N}=2$ $SU(2)$ SQCD with
four fundamental hypermultiplets, is to be considered on an extremely
squashed four-sphere. We investigate how the ellipsoid partition
function simplifies and reproduces the Liouville correlator in the
light asymptotic limit.

The organization of this article is as follows. In Section \ref{sec:lal},
we review the light asymptotic limit of Liouville theory and derive a
simple expression for four-point correlation function on sphere in this
limit. Then in Section \ref{sec:gauge} we translate this limit into the
gauge theory side, and investigate how the simplified expression for
four-point function can be reproduced from the ellipsoid partition function.
We conclude in Section \ref{sec:concl} with some speculations on how to
interpret the result in terms of 2D gauge theory.

\section{Liouville Theory in the Light Asymptotic Limit}
\label{sec:lal}

Liouville theory is a 2D CFT of a scalar field $\phi$, whose action on a
2D surface with metric $g_{ab}$ reads
\begin{equation}
 S_\text{L}=\frac1{4\pi}\int
 \sqrt{g}{\rm d}^2\sigma\left(g^{ab}\partial_a\phi\partial_b\phi +QR\phi
 +4\pi\mu e^{2b\phi}\right).
\end{equation}
Here $b$ is the Liouville coupling, $\mu$ is called the cosmological
constant, $Q=b+b^{-1}$ and $R$ is the scalar curvature. We are
interested in the correlation functions of local operators
$V_\alpha=e^{2\alpha\phi}$,
\begin{equation}
 \Big\langle\prod_iV_{\alpha_i}(\sigma_i)\Big\rangle_\mu
 ~\equiv~ \int D\phi \exp\Big(-S_\text{L}+\sum_i2\alpha_i\phi(\sigma_i)\Big).
\label{corr}
\end{equation}
Noticing that $\mu$ can be rescaled by shifting $\phi$ by constant, one
finds that the $\mu$-dependence of correlators is rather
simple. On genus-$g$ surfaces, the correlators are proportional to
$\mu^{(Q(1-g)-\sum_i\alpha_i)/b}$.

Liouville theory has an interpretation as the theory of 2D fluctuating
metric ${\rm d}s^2=e^{2b\phi}g_{ab}{\rm d}\sigma^a{\rm d}\sigma^b$,
in which classical solutions correspond to metrics with constant negative
curvature. In the presence of source, the equation of motion becomes
\begin{equation}
 \frac1{2\pi}\nabla^2\phi-\frac1{4\pi}QR-2b\mu e^{2b\phi}
 +\sum_i2\alpha_i\delta^2(\sigma-\sigma_i)=0.
\end{equation}
The role of the sources is to introduce conical deficit $\sim\alpha_i$.
By integrating this equation one finds that the area of the surface $A$ obeys
\begin{equation}
 b\mu A = \sum_i\alpha_i-Q(1-g)\,.\quad
\left(A\equiv\int\sqrt{g}{\rm d}^2\sigma e^{2b\phi}\right)
\end{equation}
As was explained in \cite{Seiberg:1990eb,Zamolodchikov:1995aa}, the
Liouville equation does not have solution if the right hand side
is negative, for example on sphere $(g=0)$ with small Liouville
momenta $(\alpha_i\sim0)$. In such cases, one needs to consider
correlation functions with fixed area $A$ by promoting $\mu$ to a
Lagrange multiplier,
\begin{equation}
  \Big\langle\prod_iV_{\alpha_i}(\sigma_i)\Big\rangle_A
 ~\equiv~  \int_{i\mathbb R}\frac{{\rm d}\mu}{2\pi i} \,e^{\mu A}
  \Big\langle\prod_iV_{\alpha_i}(\sigma_i)\Big\rangle_\mu.
\label{corrA}
\end{equation}
The classical equation of motion with the fixed-area constraint then has
solutions with constant positive curvature and negative $\mu$. The
fixed-area correlator (\ref{corrA}) has a simple power-law dependence
$\sim A^{-1+(\sum_i\alpha_i+Q(g-1))/b}$, and by Laplace transforming it
back one obtains the original correlator (\ref{corr}).

We are interested in the behavior of correlators in the limit $b\to0$.
For simplicity we will limit our discussions to correlators on
sphere, which is conformally equivalent to the complex $z$-plane. In this
case, by rescaling the Liouville field as $b\phi\equiv\varphi$ and
redefining the cosmological constant suitably, one can rewrite the
action so that $b$ appears only in an overall factor,
\begin{equation}
 S_\text{L}~=~\frac1{\pi b^2}\int{\rm d}^2z
 \Big(\partial\varphi\bar\partial\varphi + \tilde\mu\pi e^{2\varphi}\Big).
\end{equation}
Therefore, the limit $b\to0$ is a semiclassical limit. Correlators can
then be approximated by inserting into the action the classical solution
of Liouville equation with sources. Unless we consider the correlators
with heavy sources such that $\alpha_i\sim b^{-1}$, the source terms can
all be neglected. In the so-called {\it light asymptotic limit} where
all the sources satisfy
\[
\alpha_i=b\eta_i~~\text{with}~~\eta_i~~\text{fixed}, 
\]
the semiclassical approximation for correlators turns out to have a
nice integral expression, as we now review.

In the light asymptotic limit, all we need for evaluating correlators is
the classical solutions of sourceless Liouville equation, which can be
constructed as follows. Let $t$ be the stress tensor
$t\equiv-(\partial\varphi)^2+\partial^2\varphi$, and let us also
introduce $\psi\equiv e^{-\varphi}$. It follows from Liouville
equation that $\bar\partial t=0$ and $\partial^2\psi=-t\psi$. Now, if
$\varphi$ is a solution to the sourceless Liouville equation which is
smooth everywhere on sphere, then $t$ must be a holomorphic $(2,0)$
tensor which is regular everywhere on sphere, so it must vanish. $\psi$
must then be a linear function of the holomorphic coordinate $z$.
Repeating the same argument for the $\bar z$-dependence,
one finds the general sourceless solution on sphere,
\begin{equation}
 e^{-\varphi}=sz\bar z+tz+u\bar z+v
 \equiv \big(z~1\big)\,g\left(\begin{array}{c}\bar z\\1\end{array}\right),
 \quad
 g\equiv\left(\begin{array}{cc}s&t\\u&v\end{array}\right).
\end{equation}
The Liouville equation requires $\text{det}g=-\tilde\mu\pi$, which is
positive after promoting $\tilde\mu$ to a Lagrange multiplier as
explained above. For a suitable choice of the area $A$, the classical
solutions are parametrized by a $2\times2$ Hermitian matrix $g$ with
positive trace, unit determinant.
The moduli space of solutions is therefore the 3D hyperboloid $H_3^+$.

After Laplace transforming back one finds, up
to a factor proportional to $\mu^{Q/b-\sum_i\eta_i}$, the 
correlators in the light asymptotic limit is given by an integral over
$g\in H_3^+$,
\begin{equation}
 \Big\langle\prod_iV_{\alpha_i}(z_i)\Big\rangle\Big|_{b\to0} ~=~
 \int_{H_3^+}{\rm d}g\prod_i\Phi^{\eta_i}_{z_i}(g)
 ~\equiv~ \Big\langle\prod_i\Phi^{\eta_i}_{z_i}\Big\rangle_{H_3^+},
\end{equation}
where $\alpha_i=b\eta_i$ and
\begin{equation}
 \int_{H_3^+}{\rm d}g=
 \int_{\mathbb R_+}\frac{{\rm d}s}s\int_{\mathbb C}{\rm d}t{\rm d}\bar t\,,
 \quad
 \Phi^\eta_z(g)\equiv
 \left[
 \big(z~1\big)\,g\left(\begin{array}{c}\bar z\\1\end{array}\right)
 \right]^{-2\eta}\,.
\end{equation}
The explicit evaluation of the integral yields the three-point function,
\begin{eqnarray}
 \Big\langle\Phi^{\eta_1}_{z_1}\Phi^{\eta_2}_{z_2}\Phi^{\eta_3}_{z_3}
 \Big\rangle_{H_3^+}
 &=& |z_{12}|^{2\eta_{3-1-2}}|z_{23}|^{2\eta_{1-2-3}}|z_{31}|^{2\eta_{2-3-1}}
 \nonumber \\ && \cdot
 \frac\pi2\frac
 {\Gamma(\eta_{1+2+3}-1)\Gamma(\eta_{1+2-3})
  \Gamma(\eta_{2+3-1})\Gamma(\eta_{3+1-2})}
 {\Gamma(2\eta_1)\Gamma(2\eta_2)\Gamma(2\eta_3)},
\label{3pt}
\end{eqnarray}
where we used the shorthand notations $z_{12}\equiv z_1-z_2$,
$\eta_{3-1-2}\equiv\eta_3-\eta_1-\eta_2$, etc.
Two-point functions can be obtained from this by taking suitable limits,
from which one finds that $\Phi^\eta_z(g)$ with $\eta\in\frac12+i\mathbb R$
forms a complete set of normalizable wave functions on $H_3^+$. Moreover,
$\Phi^\eta_z$ and $\Phi^{1-\eta}_z$ are related via
\begin{equation}
 \Phi^\eta_z(g)=
 \frac{1-2\eta}\pi\int{\rm d}^2w|z-w|^{-4\eta}\Phi^{1-\eta}_w(g)\,.
\end{equation}
The completeness of wave functions leads to the equality
\begin{equation}
 \delta(g-g')~=~ -\frac1{2\pi^3}\int_{\frac12+i\mathbb R}\!\!\!\!{\rm d}\eta
 \int_{\mathbb C}{\rm d}^2z\;(2\eta-1)^2\Phi^\eta_z(g)\Phi^{1-\eta}_z(g')\,,
\label{cmpl}
\end{equation}
which allows us to express higher-point correlators in terms of the
three-point functions in the light asymptotic limit. In fact the space $H_3^+$
is nothing but the Wick-rotated $AdS_3$, and the wave function
$\Phi^\eta_z(g)$ is an important tool to study the CFT on that
background in the mini-superspace limit. See \cite{Teschner:1997fv} for
more detail.

As the simplest application of the above formula, let us consider the
four-point functions. Using (\ref{cmpl}) and (\ref{3pt}) one can rewrite
it as follows,
\begin{eqnarray}
 \Big\langle\Phi^{\eta_1}_{z_1}\Phi^{\eta_2}_{z_2}
 \Phi^{\eta_3}_{z_3}\Phi^{\eta_4}_{z_4}\Big\rangle_{H_3^+}
  &=&
  -\int\frac{{\rm d}\eta_0{\rm d}^2z_0}{2\pi^3}(2\eta_0-1)^2\,
 \Big\langle\Phi^{\eta_1}_{z_1}
 \Phi^{\eta_2}_{z_2}\Phi^{\eta_0}_{z_0}\Big\rangle_{H_3^+}
 \Big\langle\Phi^{1-\eta_0}_{z_0}
 \Phi^{\eta_3}_{z_3}\Phi^{\eta_4}_{z_4}\Big\rangle_{H_3^+}
 \nonumber \\ &=&
 \frac18\int{\rm d}\eta_0\,
 \frac
 {\Gamma(\eta_{1+2+0}-1)\Gamma(\eta_{1+2-0})
  \Gamma(\eta_{1-2+0})\Gamma(\eta_{2-1+0})}
 {\Gamma(2\eta_1)\Gamma(2\eta_2)\Gamma(2\eta_0-1)}
 \nonumber \\ && \hskip11mm\times
 \frac
 {\Gamma(\eta_{3+4+0}-1)\Gamma(\eta_{3+4-0})
  \Gamma(1+\eta_{3-4-0})\Gamma(1+\eta_{4-3-0})}
 {\Gamma(2\eta_3)\Gamma(2\eta_4)\Gamma(1-2\eta_0)}\times I,
 \nonumber \\
\label{4pt0}
\end{eqnarray}
with
\begin{equation}
 I \equiv
 \int\frac{{\rm d}^2z_0}\pi
 |z_{12}|^{2\eta_{0-1-2}}|z_{20}|^{2\eta_{1-2-0}}|z_{01}|^{2\eta_{2-0-1}}
 |z_{34}|^{2-2\eta_{0+3+4}}|z_{40}|^{2\eta_{3-4+0}-2}
 |z_{03}|^{2\eta_{4+0-3}-2}.
\label{4pt1}
\end{equation}
It follows from global conformal symmetry on sphere that, after some
powers of coordinate differences are factored out, the four-point
function depends on $z_1,\cdots, z_4$ only through the cross-ratio
$q\equiv z_{43}z_{21}/z_{42}z_{31}$. The $z_0$-integral (\ref{4pt1}) can
be rewritten in terms of hypergeometric functions,
\begin{eqnarray}
I &=&
 |z_{13}|^{2\eta_{-1-2-3+4}}
 |z_{14}|^{2\eta_{-1+2+3-4}}
 |z_{24}|^{-4\eta_2}
 |z_{34}|^{2\eta_{1+2-3-4}}
 \nonumber \\ &&
 \times\Bigg\{
   \frac{\Gamma(\eta_{0+3-4})\Gamma(\eta_{0+4-3})\Gamma(1-2\eta_0)}
        {\Gamma(1-\eta_{0+3-4})\Gamma(1-\eta_{0+4-3})\Gamma(2\eta_0)}
 \cdot F_1(q)F_1(\bar q)
 \nonumber \\ &&\hskip6mm
 +\frac{\Gamma(1-\eta_{0+1-2})\Gamma(1-\eta_{0+2-1})\Gamma(2\eta_0-1)}
       {\Gamma(\eta_{0+1-2})\Gamma(\eta_{0+2-1})\Gamma(2-2\eta_0)}
 \cdot F_2(q)F_2(\bar q)
 \Bigg\},
 \nonumber \\ &&
  F_1(q)\equiv q^{\eta_{0-1-2}}F(\eta_{0-1+2},\eta_{0+3-4};2\eta_0;q)\,,
 \nonumber \\ &&
  F_2(q)\equiv q^{1-\eta_{0+1+2}}F(1-\eta_{0+1-2},1-\eta_{0-3+4};2-2\eta_0;q)\,.
\end{eqnarray}
Inserting this result back to (\ref{4pt0}), one finds the expression
simplifies further due to the fact that $F_1$ and $F_2$ are related by
$\eta_0\leftrightarrow 1-\eta_0$. We finally arrive at a nice and
suggestive formula for the four-point function,
\begin{eqnarray}
\lefteqn{
 \Big\langle\Phi^{\eta_1}_{z_1}\Phi^{\eta_2}_{z_2}
 \Phi^{\eta_3}_{z_3}\Phi^{\eta_4}_{z_4}\Big\rangle_{H_3^+}
} \nonumber \\
  &=&
 \frac14
 |z_{12}|^{-2\eta_{1+2}}
 |z_{13}|^{2\eta_{-1-2-3+4}}
 |z_{14}|^{4\eta_2+2\eta_{3-4}}
 |z_{23}|^{2\eta_{1+2}}
 |z_{24}|^{-4\eta_2}
 |z_{34}|^{-2\eta_{3+4}}
 \nonumber \\ && \times
 \int{\rm d}\eta_0
 \frac
 {\Gamma(\eta_{0+1+2}-1)\Gamma(\eta_{-0+1+2})
  \Gamma(\eta_{0+1-2})\Gamma(\eta_{0-1+2})}
 {\Gamma(2\eta_0)\Gamma(2\eta_0-1)
  \Gamma(2\eta_1)\Gamma(2\eta_2)}
 \nonumber \\ && \hskip11mm\times
 \frac
 {\Gamma(\eta_{0+3+4}-1)\Gamma(\eta_{-0+3+4})
  \Gamma(\eta_{0+3-4})\Gamma(\eta_{0-3+4})}
 {\Gamma(2\eta_3)\Gamma(2\eta_4)}
 \nonumber \\ && \hskip11mm\times
 |q|^{2\eta_0}
 F(\eta_{0-1+2},\eta_{0+3-4};2\eta_0;q)
 F(\eta_{0-1+2},\eta_{0+3-4};2\eta_0;\bar q)\,.
\label{4pt2}
\end{eqnarray}
We are interested in how this expression is reproduced from the
corresponding 4D gauge theory on ellipsoids.

\section{Gauge Theory on Ellipsoids}
\label{sec:gauge}

According to the AGT relation, Liouville four-point correlators on
sphere can be reproduced from the partition function of 4D ${\cal N}=2$
SQCD with $SU(2)$ gauge group and $N_f=4$ fundamental hypermultiplets
defined on an ellipsoid with suitable background fields,
\begin{equation}
 \frac{x_0^2}{r^2}+\frac{x_1^2+x_2^2}{\ell^2}
+\frac{x_3^2+x_4^2}{\tilde\ell^2}~=~1.
\label{ell}
\end{equation}
The Liouville coupling $b$ sets the ratio of two axis-lengths as
$b=\sqrt{\ell/\tilde\ell}$, while $r$ can be chosen arbitrarily.
The limit $b\to0$ therefore corresponds to an extremely squashed $S^4$.

The exact formula for the partition function on ellipsoids was obtained using the
localization principle \cite{Pestun:2007rz, Hama:2012bg}. General
${\cal N}=2$ gauge theories on ellipsoids have a single supercharge
which squares to an element of the $U(1)\times U(1)$ isometry group. Due to
this sypersymmetry, the path integral localizes onto saddle points
labelled by an element $a$ of the Cartan subalgebra of the gauge
symmetry algebra. At the saddle point labelled by $a$, the scalar
fields in the vector multiplet take constant value proportional to $a$.
All the other fields have to vanish up to gauge equivalence, except that
the gauge field may have delta-function excitations at the north and
the south poles $(x_0,\cdots,x_4)=(\pm r,0,0,0,0)$. The exact partition
function thus takes the schematic form
\begin{equation}
 Z~=~ \int{\rm d}a\,Z_\text{cl}\cdot Z_\text{1-loop}\cdot|Z_\text{inst}|^2,
\label{Zgauge}
\end{equation}
where $Z_\text{cl}=e^{-S_\text{cl}}$ arises from the nonzero classical
value of the super Yang-Mills action, the one-loop determinant
$Z_\text{1-loop}$ arises from integrating over fluctuations
around the saddle point and $Z_\text{inst}$ \cite{Nekrasov:2002qd}
describes the contribution of instantons localized at each pole.

Various parameters of the SQCD enter the above formula, and we need their relations to the parameters of the Liouville four-point function.
First, the complexified gauge coupling is related to the cross-ratio of
the four coordinates by
\begin{equation}
 q~=~ \exp\left(-\frac{8\pi^2}{g^2}+i\theta\right)\,.
\end{equation}
The instanton partition function $Z_\text{inst}$ describing the
contributions of instantons at the north pole (or anti-instantons at the
south pole) is a power series in $q$ (resp. $\bar q$).
Next, the SQCD has mass parameters for the four hypermultiplets
$\mu_1,\cdots,\mu_4$, which are related to the Liouville momenta of the
four vertex operators $\alpha_i=\frac Q2+ip_i$ as follows,
\begin{equation}
 \mu_1=p_1+p_2,~~~
 \mu_2=-p_1+p_2,~~~
 \mu_3=-p_3+p_4,~~~
 \mu_4=-p_3-p_4.
\label{pmap}
\end{equation}
Recalling $p_i\sim \frac{iQ}2$ in the light asymptotic limit, one finds
that $\mu_1$ and $\mu_4$ become very large, while $\mu_2,\mu_3$ are small.
Similarly, the saddle point parameter $a$ is related to the
Liouville momentum $\alpha_0$ of the intermediate states (created by the
fusion of $V_{\alpha_1}$ and $V_{\alpha_2}$) as $\alpha_0=\frac Q2+ia$.

Let us now investigate each factor of the integrand of (\ref{Zgauge})
using the result of \cite{Hama:2012bg}. First, the classical value of
the Yang-Mills action for the vector multiplet at the saddle point $a$ is
$S_\text{cl}=16\pi^2 a^2/g^2$, so that $Z_\text{cl}$ becomes
\begin{equation}
 Z_\text{cl}= |q|^{2a^2} ~\stackrel{b\to0}\sim~ |q|^{-\frac1{2b^2}-1+2\eta_0}
\end{equation}
in the light asymptotic limit. This reproduces the factor
$|q|^{2\eta_0}$ in the integrand of (\ref{4pt2}).

Second, the one-loop determinant consists of the vector- and
hypermultiplet contributions
$Z_\text{1-loop}=Z^\text{(vec)}_\text{1-loop}Z^\text{(hyp)}_\text{1-loop}$,
where
\begin{eqnarray}
 Z^\text{(vec)}_\text{1-loop} &=&
 \Upsilon(2ia)\Upsilon(-2ia)~=~
 \Upsilon(2b\eta_0-Q)\Upsilon(2b\eta_0),
 \nonumber \\
 Z^\text{(hyp)}_\text{1-loop} &=&
 \prod_{i=1}^4\Upsilon(\tfrac Q2+ia+i\mu_i)^{-1}
              \Upsilon(\tfrac Q2-ia+i\mu_i)^{-1}
 \nonumber \\ &=&\Big\{
 \Upsilon(b\eta_{0+1+2}-Q)\Upsilon(b\eta_{0+1-2})
 \Upsilon(b\eta_{0-1+2})\Upsilon(b\eta_{-0+1+2})
 \nonumber \\ && \times
 \Upsilon(b\eta_{0+3+4}-Q)\Upsilon(b\eta_{0+3-4})
 \Upsilon(b\eta_{0-3+4})\Upsilon(b\eta_{-0+3+4})\Big\}^{-1}\,.
\label{Z1l}
\end{eqnarray}
The function $\Upsilon(x)$, which was first introduced in
\cite{Zamolodchikov:1995aa}, can be formally expressed as a
regularized infinite product,
\begin{equation}
 \Upsilon(x)\propto \prod_{m,n\ge0}
 \big(x+mb+nb^{-1}\big)\big(x-mb-nb^{-1}-Q\big)\,,
\end{equation}
so it was used in \cite{Hama:2012bg} to express the product of
eigenvalues in the one-loop determinant. The functions $\Upsilon(bx)$ and
$\Upsilon(bx-Q)$ appearing in (\ref{Z1l}) both have infinitely many
semi-infinite arrays of zeroes. The zeroes in each array are integer
spaced, and the arrays are mutually separated by the distance
$b^{-2}$. In the limit $b\to0$, let us approximate these functions by
keeping only the array of zeroes starting near the origin. Then
\cite{Quine:1993aa}
\begin{eqnarray}
 \Upsilon(bx) &\stackrel{b\to0}\sim &
 \prod_{m\ge0}(bx+mb)~=~ b^{\frac12-x}\frac{\sqrt{2\pi}}{\Gamma(x)},
 \nonumber \\
 \Upsilon(bx-Q) &\stackrel{b\to0}\sim &
 \prod_{m\ge0}(bx+mb-b)~=~ b^{\frac32-x}\frac{\sqrt{2\pi}}{\Gamma(x-1)}.
\end{eqnarray}
Substituting these into (\ref{Z1l}) one obtains
\begin{eqnarray}
 Z^\text{(vec)}_\text{1-loop} &\stackrel{b\to0}\sim&
 \frac{2\pi b^{2-4\eta_0}}
      {\Gamma(2\eta_0)\Gamma(2\eta_0-1)},
 \nonumber \\
 Z^\text{(hyp)}_\text{1-loop} &\stackrel{b\to0}\sim&
 \frac{b^{-6+4\eta_0+2\eta_{1+2+3+4}}}{(2\pi)^4}
 \times\Gamma(\eta_{0+1+2}-1)\Gamma(\eta_{0+1-2})
       \Gamma(\eta_{0-1+2})\Gamma(\eta_{-0+1+2})
 \nonumber \\ && \hskip30mm\times\,
       \Gamma(\eta_{0+3+4}-1)\Gamma(\eta_{0+3-4})
       \Gamma(\eta_{0-3+4})\Gamma(\eta_{-0+3+4})\,.
\end{eqnarray}
The product of Gamma functions reproduces part of the integrand of
(\ref{4pt2}), and the $\eta_0$-dependences in the powers of $b$ cancel out.
The mismatch by a product of $b^{2\eta_i}\Gamma(2\eta_i)~(i=1,\cdots,4)$
can be absorbed by renormalizing Liouville vertex operators.

Finally, let us look at the instanton series $Z_\text{inst}$. Following
the construction summarized in \cite{Alday:2009aq}, we first construct
the instanton series in $U(2)$ SQCD with two anti-fundamental hypers of
mass $\mu_1,\mu_2$ and two fundamental hypers of mass $\mu_3,\mu_4$, and
then factor out the ``$U(1)$ part''. In our problem, the Omega deformation
parameters should be chosen as $\epsilon_1=b,\epsilon_2=b^{-1}$.

In $U(2)$ theory, the various contributions of instantons localized at
the north pole are labelled by a pair of Young tableaux
$\vec Y\equiv(Y_1,Y_2)$, and the number of boxes $|\vec Y|\equiv|Y_1|+|Y_2|$
corresponds to the number of instantons. As usual, each Young tableau
$Y$ is characterized by the heights of columns
$(\lambda_1\ge\lambda_2\ge\cdots)$ or the lengths of rows
$(\lambda'_1\ge\lambda'_2\ge\cdots)$, and each box in $Y$ is labelled by
its coordinate $[m,n]\in\mathbb Z_+^2$ satisfying $1\le m\le\lambda'_n$
and $1\le n\le\lambda_m$. On the Coulomb branch vacuum labelled by the
element $\vec a=(a_1,a_2)=(a,-a)$ of Cartan subalgebra of $U(2)$, the instanton
series is given by
\begin{equation}
 Z^{U(2)}_\text{inst} ~=~
 \sum_{\vec Y}q^{|\vec Y|}\cdot z_\text{vec}(\vec a,\vec Y)
 \prod_{j=1,2}z_\text{antifund}(\vec a,\mu_j,\vec Y)
 \prod_{j=3,4}z_\text{fund}(\vec a,\mu_j,\vec Y),
\end{equation}
where\footnote{
We use the formula in the appendix of \cite{Alday:2009aq} with slight
replacements $\vec a_\text{there}=i\vec a_\text{here}$,
$m_\text{there}=\frac Q2+i\mu_\text{here}$.}
\begin{eqnarray}
 z_\text{fund}(\vec a,\mu,\vec Y) &=&
 z_\text{antifund}(\vec a,-\mu,\vec Y) \nonumber \\ &=&
 \prod_{\alpha=1}^2\prod_{[m,n]\in Y_\alpha}
 \Big(ia_\alpha+b(m-1)+\frac1b(n-1)-i\mu+\frac Q2\Big)\,,
 \nonumber \\
 z_\text{vec}(\vec a,\vec Y) &=& \prod_{\alpha,\beta=1}^2
 \prod_{[m,n]\in Y_\alpha}
 \Big(ia_\alpha-ia_\beta
      -b\{\lambda'_n(Y_\beta)-m\}+\frac1b\{\lambda_m(Y_\alpha)-n+1\}\Big)^{-1}
 \nonumber \\ && \hskip21mm\cdot
 \Big(ia_\beta-ia_\alpha
   +b\{\lambda'_n(Y_\beta)-m+1\}-\frac1b\{\lambda_m(Y_\alpha)-n\}\Big)^{-1}\,.
 \nonumber \\
\end{eqnarray}

In the above formula for the instanton series, the coefficient of
$q^{|\vec Y|}$ for each term is a ratio of polynomials in $a$ and
$\mu_k$ of order $4|\vec Y|$. The polynomials in the numerator and
denominator are the contributions from the hypermultiplets and
the vector multiplet, respectively. At first sight, they both behave as
$b^{-2\cdot 4|\vec Y|}$ in the light asymptotic limit and the ratio is
finite but, as was found in \cite{Mironov:2009qn}, their actual behavior
is more interesting.
\begin{eqnarray}
 \text{(denominator)} &\sim&
 b^{2\lambda'_1(Y_1)+\lambda'_2(Y_1)+\lambda'_1(Y_1)-8|\vec Y|},
 \nonumber \\
 \text{(numerator)} &\sim&
 b^{2\lambda'_1(Y_1)+2\lambda'_2(Y_1)+2\lambda'_1(Y_1)-8|\vec Y|}.
\end{eqnarray}
For most pairs of Young tableaux, the numerator becomes vanishingly
smaller than the denominator as $b\to0$.
The terms which remain finite in this limit are therefore labelled by $\vec Y$
for which $Y_1$ has a single row and $Y_2$ is null. The sum over pairs
of Young tableaux then simplifies to the following series
\begin{eqnarray}
 Z_\text{inst}^{U(2)} ~&\stackrel{b\to 0}{\simeq}&~
 \sum_{k\ge0}\frac{q^k}{k!}\prod_{m=0}^{k-1}
 \frac{(\eta_{0-1+2}+m)(\eta_{0+3-4}+m)}{2\eta_0+m}
 \nonumber \\ &=&
 F(\eta_{0-1+2},\eta_{0+3-4};2\eta_0;q)\,,
\end{eqnarray}
reproducing the hypergeometric function in (\ref{4pt2}). Finally, to get
the $SU(2)$ instanton sum, one has to divide the above result by the
$U(1)$ factor \cite{Alday:2009aq},
\begin{equation}
 Z^{U(1)}~=~(1-q)^{\frac12\{Q+i(\mu_1+\mu_2)\}\{Q-i(\mu_3+\mu_4)\}}
 ~=~ (1-q)^{2b^2\eta_2\eta_3}\,,
\end{equation}
which has no effect in the light asymptotic limit. 

\section{Concluding Remarks}
\label{sec:concl}

The simple formula for the Liouville four-point correlators in the
light asymptotic limit, and the corresponding simplification of the 4D
ellipsoid partition function, both suggest that a certain 2D gauge theory
may describe the limit. In particular, our investigation shows that the
sum over Young tableaux simplifies to a sum over linear arrays of boxes,
which may imply that the instanton sum turns into a vortex sum in this
limit. Note that the reduction of instanton partition functions to vortex partition functions also appears in other situation \cite{Bonelli:2011fq,Bonelli:2011wx}.

The light asymptotic limit corresponds to the limit of an extremely
squashed 4-sphere, i.e. $\ell\ll\tilde\ell$ in (\ref{ell}). In this
limit, one has various choices regarding the behavior of the other
axis-length $r$. In one typical choice $r=\ell\ll\ell$, the ellipsoid
degenerates to a small $S^2$ fibered over a large disk with a small
Omega deformation about the origin, and becomes $S^2\times \mathbb R^2$
in the limit. In another typical choice $r=\tilde\ell\gg\ell$, the role
of the base and fiber is interchanged, and one has the $(x_1,x_2)$-plane
$\mathbb R^2$ with a large Omega deformation fibered over a large
$S^2$. Our result could therefore be compared with the recent results on
SUSY gauge theories on $S^2$ \cite{Benini:2012ui,Doroud:2012xw} or
squashed $S^2$ \cite{Gomis:2012wy},
especially the theories appearing on 2D surface defects corresponding to
Liouville degenerate operator insertions
\cite{Dimofte:2010tz,Dorey:2011pa,Chen:2011sj,Doroud:2012xw}.
At present, it is not clear to us which is the more suitable
picture to understand the simplification of 4D partition function. In
either picture, since $\epsilon_1/\epsilon_2\to 0$ in our limit, the
physics near the north and south poles of the ellipsoid is a 2D
${\cal N}=(2,2)$ SUSY theory in the Nekrasov-Shatashvili limit
\cite{Nekrasov:2009rc}. Furthermore, in addition to this ``dimensional
reduction'', one also has to send some of the mass parameters to
infinity to reproduce the Liouville light asymptotic limit. It would be
interesting to understand fully the mechanism of this dimensional
reduction.

\paragraph{Acknowledgment.}

KH thanks Heng-Yu Chen and Sungjay Lee for discussions, and also
thank the Simons Center for Geometry and Physics for hospitality during
the completion of this work.
NH also thanks Tomoki Nosaka for discussions.
The authors are particularly thankful to Sylvain Ribault for participation in the 
early stages of this project.
We are grateful to the hospitality of the colleagues in IPhT, CEA
Saclay during their stay. This research has received funding from the [European
Union] Seventh Framework Programme [FP7-People-2010-IRSES] under grant
agreement No. 269217, the PHC SAKURA 2012, Projet No. 27588UASakura and
the corresponding Grant from Japan. The work of NH is also supported in
part by the JSPS Research Fellowships for Young Scientists.

\end{document}